\documentclass[12pt]{iopart}
\usepackage{graphicx}  
\usepackage[T1]{url}   

\begin{document}
\title[A Bayesian method to set upper limits on the strength of a periodic
gravitational wave]{A Bayesian method to set upper limits on the strength of a periodic
gravitational wave signal from the remnant of SN1987A: possible applications in LIGO searches.}
\author{Richard Umst\"atter$^1$,
Renate Meyer$^1$,
Nelson Christensen$^2$}
\address{$^1$Department of Statistics, University of Auckland,
Auckland, New Zealand\\
$^2$Physics and Astronomy, Carleton College,
Northfield, MN 55057, USA\\}
\begin{abstract}
We present a method that assesses the theoretical detection limit of a Bayesian Markov chain Monte Carlo search for a periodic gravitational wave signal emitted by a neutron star. Inverse probability yields an upper limit estimate for the strength when a signal could not be detected in an observed data set. The proposed method is based on Bayesian model comparison that automatically quantifies Occam's Razor. It limits the complexity of a model by favoring the most parsimonious model that explains the data. By comparing the model with a signal from a pulsar to the null model that assumes solely noise, we derive 
the detection probability and an estimate for the upper limit that a search, for example, for a narrow-band emission for SN1987a, might yield on data at the sensitivity of LIGO data for an observation time of one year. 
\end{abstract}

\newcommand{\avec}{\mbox{\boldmath$a$}}
\newcommand{\pvec}{\mbox{\boldmath$a$}}
\newcommand{\avecs}{\mbox{\scriptsize\boldmath$a$}}
\newcommand{\bvec}{\mbox{\boldmath$b$}}
\newcommand{\dvec}{\mbox{\boldmath$d$}}
\newcommand{\dvecs}{\mbox{\scriptsize\boldmath$d$}}
\newcommand{\fvec}{\mbox{\boldmath$f$}}
\newcommand{\hvec}{\mbox{\boldmath$d$}}
\newcommand{\hvecs}{\mbox{\scriptsize\boldmath$d$}}
\newcommand{\rvec}{\mbox{\boldmath$r$}}
\newcommand{\tvec}{\mbox{\boldmath$t$}}
\newcommand{\aalpha}{\mbox{\boldmath$\alpha$}}
\newcommand{\se}{{\rm se}}
\newcommand{\lh}{l_{h_0}}

\newcommand{\be}{\begin{equation}}
\newcommand{\ee}{\end{equation}}

\newcommand{\rd}{\,{\rm d}}
\newcommand{\D}{\mbox{\boldmath$D$}}
\renewcommand{\b}{\mbox{\boldmath$b$}}
\newcommand{\bepsilon}{\mbox{\boldmath$\epsilon$}}
\renewcommand{\vec}[1]{\mathbf{#1}}
\newcommand{\AFkp}{A^{+}F_k^{+}}
\newcommand{\AFkc}{A^{\times}F_k^{\times}}
\newcommand{\sinPSI}{\sin(\Delta\Phi_k)}
\newcommand{\cosPSI}{\cos(\Delta\Phi_k)}
\newcommand{\sinSQPSI}{\sin^2(\Delta\Phi_k)}
\newcommand{\cosSQPSI}{\cos^2(\Delta\Phi_k)}
\newcommand{\sintwoPSI}{\sin(2\Delta\Phi_k)}

\newcommand{\sig}{\sigma_{k}}

\newcommand{\epsre}{\epsilon_{k,\mathrm{re}}}
\newcommand{\epsim}{\epsilon_{k,\mathrm{im}}}

\newcommand{\penal}{\mathcal{P}}

\newcommand{\cov}{{\rm Cov}}
\newcommand{\var}{{\rm Var}}
\newcommand{\E}{{\rm E}}
\newcommand{\BIC}{{\rm BIC}}
\newcommand{\DIC}{{\rm DIC}}
\newcommand{\BICdaM}[1]{\BIC_{\datas,\avecs_{*},\mathcal{M}_{#1}}}
\newcommand{\dBICda}{\Delta \BIC_{\datas,\avecs_{*}}}
\newcommand{\dBICa}{\Delta \BIC|\avec_{*}}
\newcommand{\dBICap}{\Delta \BIC|\avecp{}}
\newcommand{\VARa}{\sigma^2_{\avecs_{*}}}
\newcommand{\MEANa}{\mu_{\avecs_{*}}}

\newcommand{\VARBICa}{\sigma^2_{\BIC,\avecps}}
\newcommand{\MEANBICa}{\mu_{\BIC,\avecps}}

\newcommand{\Prob}{p}
\newcommand{\OP}{{\rm OP}}
\newcommand{\SSQ}{{\rm SSQ}}
\newcommand{\SNR}{{\rm SNR}}
\newcommand{\data}{\dvec_{*}}
\newcommand{\datas}{\dvecs_{*}}
\newcommand{\avecp}[1]{\avec^{#1}_{\bullet}}
\newcommand{\avecps}{\avecs_{\bullet}}
\newcommand{\wurzelterm}{\sqrt{a_1^2-a_2^2}}
\newcommand{\N}{{\rm N}}
\newcommand{\LH}{H_{\Sigma}}
\newcommand{\IG}{\mathcal IG}
\newcommand{\ML}{{\rm ML}}
\newcommand{\UL}{{\rm UL}}
\newcommand{\sigmavec}{\mbox{\boldmath$\sigma$}}



\section{\bf Introduction}

Several mechanisms have been proposed that would cause rapidly rotating neutron stars to emit quasi-periodic gravitational waves~\cite{Cutler:2002,Bildsten:1998}. Interferometric gravitational wave detectors that are now operating in
numerous locations around the world~\cite{GEO:1997, LIGO:1997, VIRGO:1997, TAMA:1997} now allow for their verification and much work has gone into the development of dedicated search algorithms for these signals. Radio observations can provide the sky location, rotation frequency and spin-down rate of known pulsars. 
The frequency of the reported remnant of SN1987a for example is not known accurately \cite{Middleditch:2000} but Markov Chain Monte Carlo (MCMC) methods \cite{ChristensenMeyerLibson:2004, Christensen:2004,  UMDVWC:2004, MaxEnt:2004, Veitch:2005} are able to search a range of frequencies (and other physical parameters) in a reasonable time.

As in previous studies \cite{LIGO-CW:2004, DupuisWoan:2005}
the signal under consideration is one that is expected from a non-precessing triaxial
neutron star. The gravitational wave signal from such an object is at
$f=2 f_{\rm r}$ twice its rotation frequency $f_r$, and we
characterize the amplitudes of each polarization with overall
strain factor, $h_0$. The measured gravitational wave signal will
also depend on the antenna patterns of the detector for the
`cross' and `plus' polarizations, $F_{\times, +}(t;\psi,\alpha,\delta)$, giving a signal
$s(t) \!=\!   F_{+}(t;\ldots)h_{0}(1 + \cos^{2}\iota)\!\cos
\Phi(t;\ldots)/2 + F_{\times} (t;\ldots)h_{0}\cos \iota \sin  \Phi(t;\ldots)$ where $\iota$ is the inclination angle. 
The antenna pattern of the detector depends on time $t$, the polarization angle $\psi$ and its location determined by right ascension $\alpha$ and declination angle $\delta$. The location is assumed to be known from, for example, radio observations. A simple slowdown model \cite{Jaranowski:1998} provides the phase evolution of the signal as
\be
 \Phi(t;\vec{n},f_s,\dot{f}_s) = \phi_{0} + 2\pi \left[f_{\rm s}(T_{(\alpha,\delta)} - T_{0}) + \dot{f_{\rm s}} (T_{(\alpha,\delta)}-T_{0})^{2} /2 \right],
\label{phaseevolution}
\ee
where
\be
T_{(\alpha,\delta)} = t + \delta t= t + \frac{\vec{r} \cdot \vec{n}}{c}  +
\Delta{T}
\label{TimeCorrection}
\ee
is the time of arrival of the signal at the solar system
barycenter when $t$ is the time at the detector.
Here, $\phi_{0}$ is the phase of the signal at a fiducial
time $T_{0}$, $\vec{r}$ is the position of the detector with
respect to the solar system barycenter, $\vec{n}$ is a unit vector
in the direction of the neutron star (depending on $\alpha$ and $\delta$), $c$ is the speed of light and $\Delta{T}$ contains the relativistic corrections to the
arrival time \cite{Taylor:1994}.

If $f_{\rm s}$, $\dot{f}_{\rm s}$, and $\vec{n}$ are known from
radio observations, for instance, the signal can be \emph{heterodyned} by multiplying
the data by $\exp[-i\Phi(t;\vec{n},f_s,\dot{f}_s)]$, low-pass filtered and
resampled, so that the only time varying quantity remaining is the
antenna pattern of the interferometer.
The reference sky location is also needed
for the heterodyning process prior to the MCMC simulation.
We are left with a simple model with four unknown parameters
$h_0$, $\psi$, $\phi_{0}$, and $\iota$.
If there is an uncertainty in the frequency and
frequency derivative two additional parameters come into play, the
differences between the signal and heterodyne frequency and
frequency derivatives, $\Delta f$ and $\Delta\dot{f}$. 
The unit vector $\vec{n}$ points to the right ascension $\alpha$ and declination $\delta$ of the purported neutron star. 

A detailed description of the heterodyning procedure is presented
elsewhere~\cite{LIGO-CW:2004,DupuisWoan:2005}. The model of the heterodyned signal of a pulsar has form~\cite{DupuisWoan:2005}
\begin{eqnarray}
y(t_k;\avec) &=&  F_+(t_k;\psi,\alpha,\delta)h_{0} (1 +
 \cos^2\iota)e^{i\Delta\Phi(t_k;\alpha,\delta,\Delta f,\Delta\dot{f})}\!/4 \nonumber \\
 &-&  \!i F_\times(t_k;\psi,\alpha,\delta) h_{0} \cos\iota e^{i\Delta\Phi(t_k;\alpha,\delta,\Delta f,\Delta\dot{f})}\!/2,\label{yeqPulsar} 
\end{eqnarray}
where $t_k$ is the time of the $k^{\rm th}$ bin and $\avec =
(h_0,\,\cos\iota,\,\phi_0,\,\psi,\,\Delta f,\,\Delta\dot{f})$ is a
vector of the unknown parameters. $\Delta\Phi(t;\alpha,\delta,\Delta f,\Delta\dot{f})$ represents the
residual phase evolution of the signal, equaling
$\phi_0+2\pi[\Delta f(T_{(\alpha,\delta)}-T_0)+\Delta\dot{f}(T_{(\alpha,\delta)}-T_0)^{2}/2]$, where $T_{(\alpha,\delta)}$ (Eq.~(\ref{TimeCorrection})) depends on the known sky location of the pulsar. Note, that the gravitational wave oscillates at twice the rotation frequency of the pulsar's rotation frequency. Therefore, the frequency in Eq.~\ref{yeqPulsar} refers to the gravitational wave frequency. The objective is to fit this model to the data $B_k=y(t_k;\avec) +\epsilon_k$,where~$\epsilon_k$ is assumed to be normally distributed noise
with a mean of zero and known variance $\sigma_k^2$. Assuming
statistical independence of the binned data points, $B_k$, the joint
likelihood that these data $\hvec =\{B_k\}$ arise from a model
with a certain parameter vector $\avec$ is \cite{DupuisWoan:2005}
\be
 \Prob(\hvec|{\avec})\propto\prod_k \exp\left[-\left| (B_k-y(t_k;{\avec}))/\sigma_k\right| ^2 \right]/2
   = \exp \left[-\chi^2(\avec)/2\right],
\label{jointLH}
\ee
where
\be
 \chi^2(\avec)=\sum_k{|B_k-y(t_k;\avec)|^2/\sigma^2_k}.
\label{chisquare}
\ee

In order to draw any inference on the unknown parameter vector
$\avec$ the posterior probability of $\avec$ given
$\hvec$ is needed, which can be obtained from the likelihood via an
application of Bayes' theorem. The unnormalized posterior density
$\Prob(\avec|\hvec)\propto \Prob(\avec)\Prob(\hvec|\avec)$ is the product of the prior density of $\avec$, $\Prob(\avec)$, and
the joint likelihood, $\Prob(\hvec|\avec)$. 
In this study uniform priors distributions are used with prior ranges $[0,2\pi]$, $[-\pi/4,\pi/4]$ and
$[-1,1]$ for the angle parameters $\phi_0$, $\psi$ and $\cos
\iota$ respectively.

For $h_0$, a uniform prior is specified
with boundary $[0,10^{-20}]$. For the frequency and spin down uncertainty, suitable uniform priors are used with ranges of
$[-\frac{1}{120},\frac{1}{120}]$\,Hz  and
$[-10^{-9},10^{-9}]$\,Hz\,s$^{-1}$ for $\Delta f$ and
$\Delta\dot{f}$, respectively, as applied in \cite{UMDVWC:2004}.
The normalized posterior density $\Prob(\avec|\hvec)=
\Prob(\avec)\Prob(\hvec|\avec)/\Prob(\hvec)$ cannot be evaluated analytically, therefore
Monte Carlo methods are used here to explore $\Prob(\avec|\hvec)$, as described in
\cite{UMDVWC:2004}.

When the signal-to-noise ratio (SNR) and hence the signal's evidence declines, it becomes increasingly difficult to sample efficiently from the posterior distribution using MCMC. The major problem lies in the
frequency parameters $\Delta f$ and $\Delta\dot{f}$. Long
integration periods yield narrow posterior modes and when the SNR
is small, their occurrence is also negligible with most of the
posterior probability mass spread over the entire parameter space
determined by the prior distribution. The sampling process of an
MCMC sampler becomes inefficient in covering that part of the parameter 
space where the signal is concentrated. The
question that will be addressed in this paper is the threshold of
the SNR for which MCMC sampling becomes ineffective and below which no signal parameters can be retrieved.

\section{\bf The detection of weak signals}
\label{SigInData}

The presence of a signal within the data can be assessed by a formal Bayesian model comparison of the model the contains a signal with the null model that contains no signal. Bayes factors could be applied but they require a properly converged MCMC output. Without the need of MCMC samples, this paper aims to give theoretical detection probabilities dependent on signal-to-noise ratios. 

\subsection{\bf Derivation of a theoretical detection probability}
\label{theorlimit}

For the Bayesian Information Criterion (BIC), also called the
Schwarz criterion, there is no particular need for the MCMC output
samples. The BIC is defined \cite{Kass:1995}  as \be \BIC=-2
\log(\mbox{maximum likelihood}) + \penal, \label{BICPulsar} \ee
where the penalty term $\penal=d\log n$ brings in the number of
$d=6$ independent parameters that describe the model, and the
number $n$ of data samples. The penalty term penalizes the number
of parameters in a model in order give preference to simpler
models and meet the principle of Occam's Razor.

The objective is to derive a theoretical limit for the detection of
a signal within a data set observed during a determined
observation period at a certain noise level. 
This section is dedicated to find a distribution of the BIC depending on the noise, conditioned on the parameters of a potential pulsar.

The observation period is a vector $\OP=(t_1,\ldots,t_n)'$ of $n$ time points
$t_k$ with $k\in \{1,\ldots,n\}$ during which the data has been
collected starting from $t_{\rm start}$ and ending at $t_{\rm
end}$. The noise vector is a vector
$\sigmavec=(\sigma_1,\ldots,\sigma_n)'$ for the $n$ data bins.
Given the \emph{true} parameter vector of the pulsar from which
the signal arises, the full information needed for a detection is
determined by the vector $\avec_{*}=(h^{*}_0, \cos\iota^{*},
\psi^{*},\alpha^{*}, \delta^{*}, \Delta f^{*}, \Delta
\dot{f}^{*},\sigmavec,\OP)'$. Although some parameters like the sky
location are expected to be known, they are essential factors for
the detection probability in connection with the observation
period and the noise. These are essential parts of the parameter
vector as the detection depends significantly on them.

A signal \emph{detection} depends on the actual evidence 
of the model that assumes the presence of a signal from a pulsar
within an arbitrary data set when compared to the null model of mere noise.
Each potential data set under
consideration is based on the \emph{true} parameters of a
potential pulsar. Therefore each model comparison is conditioned
on a data set $\data$ that is conditioned on the parameter vector
$\avec_{*}$. This fact can be used to obtain, for large
sample sizes, an approximation for
the maximum likelihood value since the maximum likelihood estimate
(MLE) is asymptotically consistent and efficient under certain
regularity conditions that are generally satisfied \cite{Casella:2002}. 
Thus the estimates converge to the true values for large
samples sizes. The sample sizes that we expect are in fact in the
range of tens of thousands.

A potential data set $\data$ from a pulsar, based on a \emph{true}
parameter vector $\avec_{*}$ is modeled by $\mathcal{M}_{*}:
\data^{(k)}=y(t_k;\avec_{*})+\epsilon_k$ with noise vector
$\epsilon_k$. Due to the fact that $\data$ is conditioned on
$\avec_{*}$, an approximate maximum log-likelihood under model
$\mathcal{M}_{1}$ is 
\be \log
\ML_{\datas,\avecs_{*},\mathcal{M}_{1}} \approx -
\chi^2_{\datas,\avecs_{*},\mathcal{M}_{1}}(\avec_{*})/2= -\sum_k{
\frac{\left|\epsilon_k\right|^2}{2\sigma^2_k}}, \label{MLE1} \ee
This term comprises the sum of the squared residuals as the model is
fitted by the \emph{true} parameter vector. On the other hand,
under model $\mathcal{M}_{0}$ that encompasses no parameters, the
log-likelihood has a constant value and therefore its maximum is 
\be 
\log \ML_{\datas,\avecs_{*},\mathcal{M}_{0}}=
-\chi^2_{\datas,\avecs_{*},\mathcal{M}_{0}}/2= -\sum_k{
\frac{\left|y_k(t_k;\avecs_{*})+\epsilon_k\right|^2}{2\sigma^2_k}},
\label{MLE0} 
\ee 
where the summation term contains the true and
given parameter vector of the signal. It is clear that $\log
\ML_{\datas,\avecs_{*},\mathcal{M}_{1}} \ge \log
\ML_{\datas,\avecs_{*},\mathcal{M}_{0}} \forall \avec_{*}$. As a
result of this, naturally model $\mathcal{M}_{1}$ has to be
preferred at all times. This, however, does not take into account
the penalty term that comes into play due to the principle of
Occam's razor. Equality of Eq.~\ref{MLE0} and \ref{MLE1} can only
be achieved for a zero amplitude $h^{*}_0$ in parameter vector
$\avec_{*}$. But how large do we have to choose this amplitude,
also considering other influential parameters, in order to justify
model $\mathcal{M}_{1}$ with its many more parameters? This is the
essential idea behind this model comparison approach and the
penalty terms play a key role in it.

We aim to compare model $\mathcal{M}_0$ and $\mathcal{M}_1$ conditioned on the data set $\data$, conditioned on a potential pulsar characterized by the \emph{true} parameter vector $\avec_{*}$. By substituting Eq.~(\ref{MLE1}) and Eq.~(\ref{MLE0}) into Eq.~(\ref{BICPulsar}), we obtain
\be
\BICdaM{0}=-2 \log \ML_{\datas,\avecs_{*},\mathcal{M}_{0}}
\label{BICPulsarM0}
\ee
as model $\mathcal{M}_{0}$ has $d=0$ parameters and
\be
\BICdaM{1} = -2 \log \ML_{\datas,\avecs_{*},\mathcal{M}_{1}} + \penal.
\label{BICPulsarM1}
\ee
With respect to $\mathcal{M}_0$ and $\mathcal{M}_1$, a probability for model $\mathcal{M}_1$ can be derived by
\begin{eqnarray}
\Prob(\mathcal{M}_1|\data,\avec_{*})&=&
\left(1+e^{\dBICda/2 -\log \Prob(\mathcal{M}_1)+\log \Prob(\mathcal{M}_0) }\right)^{-1}
\end{eqnarray}
Here, $\Prob(\mathcal{M}_0)$ and $\Prob(\mathcal{M}_1)$ are prior
probabilities for $\mathcal{M}_0$ and $\mathcal{M}_1$
respectively. The interested reader is referred \cite{UmstaetterThesis:2006} for a more detailed derivation. We will address different prior scenarios later but
for now, we choose equal probabilities $\Prob(\mathcal{M}_0)=\Prob(\mathcal{M}_1)=0.5$  for the models as a natural
choice when there is no prior information about the possible existence of a signal.
This yields $\Prob(\mathcal{M}_1|\data,\avec_{*})=\left(1+e^{\dBICda/2}\right)^{-1}$
where $\dBICda:=\BICdaM{1}-\BICdaM{0}$. It represents the
probability that the data $\data$ from a potential pulsar with
given parameter vector $\avec_{*}$ is better modeled by
$\mathcal{M}_1$ (a signal) rather than $\mathcal{M}_0$ (no
signal). In other words it is the probability for the existence of
a signal in the data that is emitted by a pulsar with parameter
vector $\avec_{*}$. It is merely the difference of the two BIC
values under consideration that is responsible for a signal
detection. A difference of zero for example would yield a $50$\%
probability for both models. A probability conditioned on data
$\data$ from the vector $\avec_{*}$, can be expressed as
\begin{eqnarray}
\Prob(\mathcal{M}_1|\avec_{*})=\E\left[\Prob(\mathcal{M}_1|\data,\avec_{*})|\avec_{*}\right]
=\E \left[ ( 1+e^{\dBICda/2})^{-1}|\avec_{*} \right].
\label{EBICprob}
\end{eqnarray}

There is no simple way to solve this expression analytically and
although feasible, a Monte Carlo sampling process would be 
lengthly. From a physical perspective, phase $\phi_0$
and the frequency parameters $\Delta f$, $\Delta \dot{f}$ should
have no impact on the actual signal detection as the SNR mainly
depends on the amplitude $h^{*}_0$, inclination $\cos\iota^{*}$,
noise $\sigmavec$, and observation time $\OP$. To a smaller extent
the SNR is also influenced by the course of the antenna pattern
over the observation time $\OP$ with parameters $\psi^{*}$,
$\alpha^{*}$, and $\delta^{*}$. We assume the sky location to be
known and condition on $\alpha^{*}$ and $\delta^{*}$.

The probability $\Prob(\mathcal{M}_1|\avec_{*})$ is determined by the distribution
of $\dBICda$. Thus the characteristics of
$\dBICda$ will be derived below. By using equations Eq.~(\ref{MLE1}),
Eq.~(\ref{MLE0}), Eq.~(\ref{BICPulsarM0}),
Eq.~(\ref{BICPulsarM1}) we obtain
\begin{equation}
\dBICda\approx  \sum_k{ \left|\epsilon_k\right|^2/\sigma^2_k} + \penal -\sum_k{ \left|y_k(\avecs_{*})+\epsilon_k\right|^2/\sigma^2_k}
\label{DeltaBIC1}
\end{equation}

In \cite{DupuisWoan:2005}, white Gaussian noise $\epsre,\epsim
\sim N(0,\sigma_k^2)$ is assumed where the $\sigma_k^2$ are
estimated for each bin from the noise floor in a 4 Hz band of data
around the signal frequency. By substituting $y(t_k;\avec_{*})$ of
Eq.~\ref{yeqPulsar} and defining some abbreviations, $F_{+,\times}(t_k;\psi^{*},\alpha^{*}, \delta^{*})\!:=\!F_k^{+,\times}$,
$e^{i\Delta\Phi(t_k;\alpha^{*}, \delta^{*}, \Delta f^{*}, \Delta
\dot{f}^{*})}:=e^{i\Delta\Phi_k}=\cosPSI+i\sin(\Delta\Phi_k)$,
$\frac{1}{4} h^{*}_{0}(1+\cos^2 \iota^{*})=:A^{+}$, and
$\frac{1}{2} h^{*}_{0} \cos \iota^{*}=:A^{\times}$ we can
rewrite Eq.~(\ref{DeltaBIC1}) as
\begin{eqnarray}
&&\dBICda \approx 
\penal-\sum_k \left[ (A^{+}F^{+}_k)^2/\sig^2 + (A^{\times}F^{\times}_k)^2/\sig^2 \right] \nonumber \\
&& - 2\sum_k \left( \left[A^{+}F^{+}_k \cosPSI + A^{\times}F^{\times}_k \sinPSI\right]/\sig \right) \epsre/\sig \nonumber \\
&& - 2\sum_k \left( \left[A^{+}F^{+}_k \sinPSI  -  A^{\times}F^{\times}_k \cosPSI\right]/\sig \right) \epsim/\sig. 
\label{bigsum}
\end{eqnarray}
The quadratic noise terms cancel out and we are left with
normally distributed terms. Given a pulsar with parameter vector
$\avec_{*}$, the $\dBICda$ is thus normally distributed. The terms
that contain the phase evolution canceled out as well and
Eq.~(\ref{bigsum}) is thus independent of the parameters $\phi_0$,
$\Delta f$, and $\Delta \dot{f}$. With $\epsre,\epsim \sim
N(0,\sigma^2_k)$ we have
$\E(\epsre/\sigma_k)=\E(\epsim/\sigma_k)=0$ and the expected value
of Eq.~(\ref{bigsum}) has the form
\begin{equation}
\MEANa:=\E(\dBICda) = \penal-\sum_k \sig^{-2} \left[
(A^{+}F^{+}_k)^2 + (A^{\times}F^{\times}_k)^2 \right].
\label{expectedchisq0}
\end{equation}

Eq.~(\ref{expectedchisq0}) yet allows some insight as it tells us that for a given arbitrary parameter vector $\avec_{*}$, model $\mathcal{M}_0$
would be preferred over $\mathcal{M}_1$, if $\MEANa>0$. Given a
parameter vector $\avec_{*}$, the variance of Eq.~(\ref{bigsum}) is
\begin{equation}
\VARa \!=\! \var (\dBICda) = 4 (\penal -\MEANa).
\label{variancechisq0}
\end{equation}

Both expressions Eq.~(\ref{expectedchisq0}) and Eq.~(\ref{variancechisq0}) only depend on the five parameters $h^{*}_0$, $\cos \iota^{*}$, $\psi^{*}$, $\alpha^{*}$, and $\delta^{*}$.
The parameters $\psi^{*}$, $\alpha^{*}$, and $\delta^{*}$ only
enter in the plus and cross polarization terms $F^{+}_k$ and $F^{\times}_k$ of the antenna pattern which depends on the orientation sweep of the interferometer towards the pulsar and the polarization angle of the gravitational wave that it emits. 

We are left with the random variable $\dBICa \sim
\N(\MEANa,\VARa)$ that depends on five parameters of the pulsar
plus noise $\sigmavec$ and observation period $\OP$. If we assume
constant noise $\sigma$ over time, we can combine $h^{*}_0$ and $\sigma$ to a more handy SNR
$h^{*}_0/\sigma$ parameter. We define a new vector $\avecp{}=(h^{*}_0/\sigma,
\cos \iota^{*}, \psi^{*}, \alpha^{*}, \delta^{*}, \OP)'$ with
observation period $\OP=(t_1,\ldots,t_n)'$. Explicitly, the
difference in the BIC values with respect to models
$\mathcal{M}_0$ and $\mathcal{M}_1$, for arbitrary data sets,
conditioned on $\avecp{}$ follow the distribution $\dBICap \sim
\N(\MEANBICa,\VARBICa)$ with
\begin{equation}
\MEANBICa\!=\!\penal\!\!-\!\!\left( \frac{h^{*}_{0}}{\sigma} \right)^2\!\! \left( \left[\frac{1}{4} (1+\cos^2 \iota^{*})\right]^2\!\! \sum_k \left(F_k^{+}\right)^2
 \!\!\!+\! \left[\frac{1}{2} \cos \iota^{*}\right]^2 \!\! \sum_k \left(F_k^{\times}\right)^2 \right)
\label{expectedchisq}
\end{equation}
and
\begin{equation}
\VARBICa= 4 \left(\penal -\MEANBICa\right).
\label{variancechisq}
\end{equation}
Using these information, Monte Carlo methods can be used to estimate Eq.~(\ref{EBICprob}).

As an example, we consider a data set that we can expect was taken over one year at the three LIGO interferometers Hanford (4km, 2km) and Livingston (4km) with three different noise levels at the three interferometers. 
A sensible heterodyning frequency for the SN1987a remnant is $f_{\rm s}=2 f_{\rm r}=935$Hz \cite{Middleditch:2000}.
For the purpose of illustrating an example we will assume noise levels
that are likely to be close to LIGO's S5 values at the frequency in
question. We therefore assume noise levels $8 \times 10^{-24}$ (Hanford 4km), $1.5 \times 10^{-23}$ (Hanford 2km), and $9 \times 10^{-24}$ (Livingston 4km) at that frequency and an observation period $\OP$ of one year of S5 data that would be heterodyned to a potential source at $\alpha^{*}=5^{\rm h}\,35^{\rm m}\,28.03^{\rm s}$ and $\delta^{*}=-69^\circ\,16'\,11.79''$ (SN1987a) with 525600 bins at one sample per minute. The data are analyzed for each interferometer separately and also combined by the sum of the log-likelihoods, as we assume independence.
The parameter vector encompasses $\avecp{}\!=\!\left(
h^{*}_0/\sigma,
\cos \iota^{*},
\psi^{*},
\alpha^{*}\!=\!5^{\rm h}\,35^{\rm m}\,28.03^{\rm s},
\delta^{*}\!=\!-69^\circ\,16'\,11.79'',
\rm{OP}
\right)$
in which the values of $h^{*}_0/\sigma$ and $\cos \iota^{*}$ and $\psi^{*}$ are unknown. In order to derive a probability conditioned on $h^{*}_0/\sigma$, we need to marginalize $\Prob(\mathcal{M}_1|\avecp{})$ over $\cos \iota^{*}$ and $\psi$ and obtain
$\Prob(\mathcal{M}_1|h^{*}_0/\sigma,\alpha^{*},\delta^{*},\OP)=\int{\Prob(\mathcal{M}_1|\avecp{})}
d\Prob_{\cos\iota}(\cos \iota^{*}) d\Prob_{\psi}(\psi^{*})$.

\begin{figure}[!ht]
  \begin{center}
    \begin{minipage}[t]{4.5cm}
          \includegraphics[width=4.5cm,angle=270]{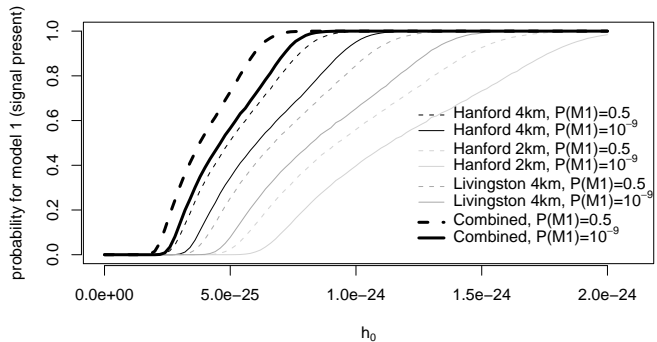}
    \end{minipage}\hfill
    \begin{minipage}[t]{8.5cm}
      \caption{Signal detection probability for the three different interferometers and two different prior probabilities for model $\mathcal{M}_1$ as a function of the amplitude $h^{*}_0$ for one year of S5 data. The curves of two prior probabilities $\Prob(\mathcal{M}_1)=0.5$ (dashed lines) and $\Prob(\mathcal{M}_1)=10^{-9}$ (solid lines) are shown.}
\label{sigdetectBIC}
    \end{minipage}
  \end{center}
\end{figure}

Fig.~\ref{sigdetectBIC} displays the probability of a signal detection as a function of the amplitude. Two different prior probabilities on the signal existence are chosen. The natural choice is $\Prob(\mathcal{M}_1)=0.5$ when there is no information available. However, we know that we focus only a $1/60$Hz band and the probability of an existence needs to be split on the frequency bands in which we expect a signal. In addition, we do not know whether there is a neutron star at all which lowers the probability further. For this reason, we chose a rather arbitrary and extremely small probability of $\Prob(\mathcal{M}_1)=10^{-9}$ in order to asses the impact of that prior probability.
We obtain the graph shown in Fig.~\ref{sigdetectBIC}.

A larger amplitude $h^{*}_0$ is required for a successful detection when we doubt the existence of a signal. Hence, the data must speak more clearly for a signal in order to overcome the low prior probability but since the observation period of one year is rather long, the effect of the prior probability is fairly small.

All graphs compiled so far are showing a signal detection probability given a particular scenario but the question we aim to answer in the next section is how strong a signal still can be even if a signal can't be seen. 

\subsection{\bf Performance of the Bayesian MCMC search in setting an upper limit using S5 data}

\begin{figure}[!ht]
  \begin{center}
    \includegraphics[width=0.48\textwidth]{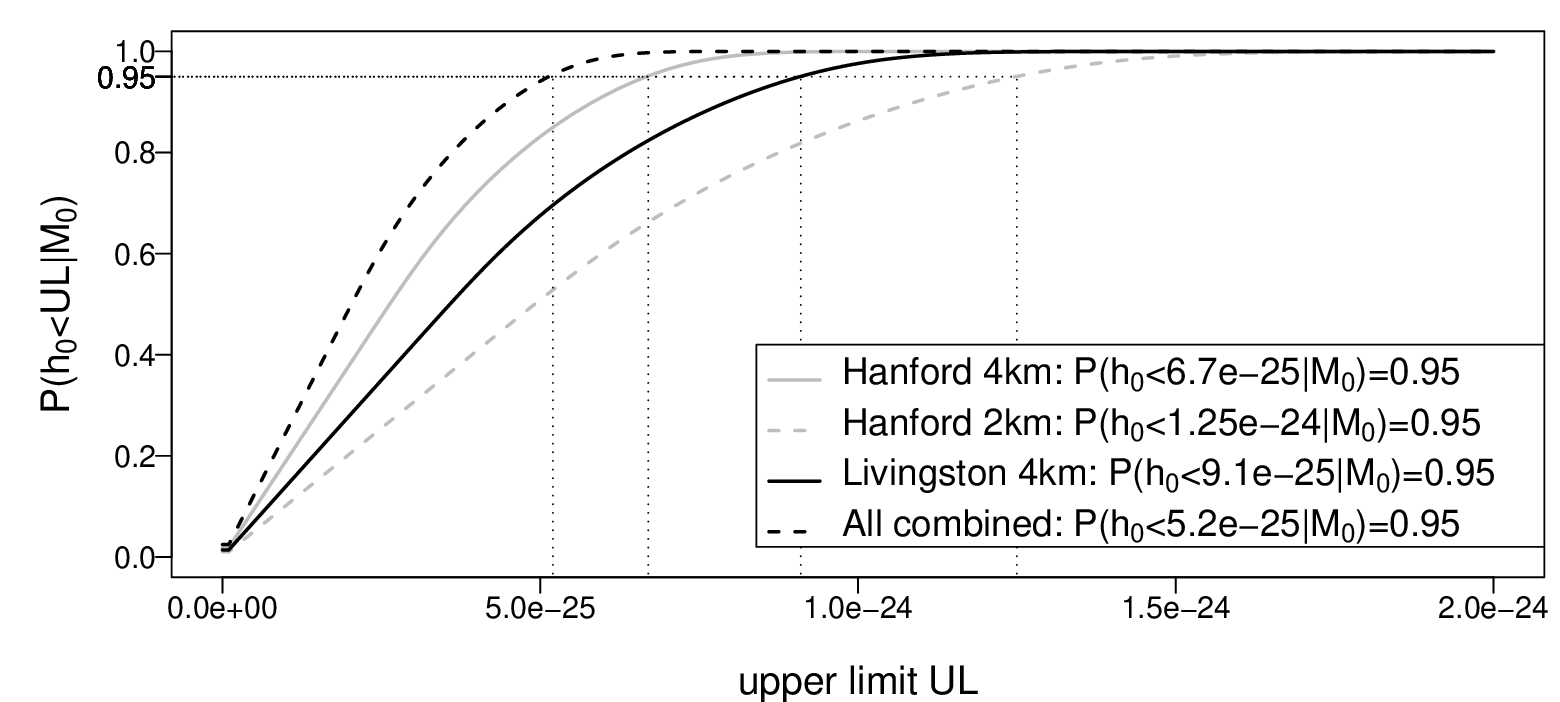}
    \includegraphics[width=0.48\textwidth]{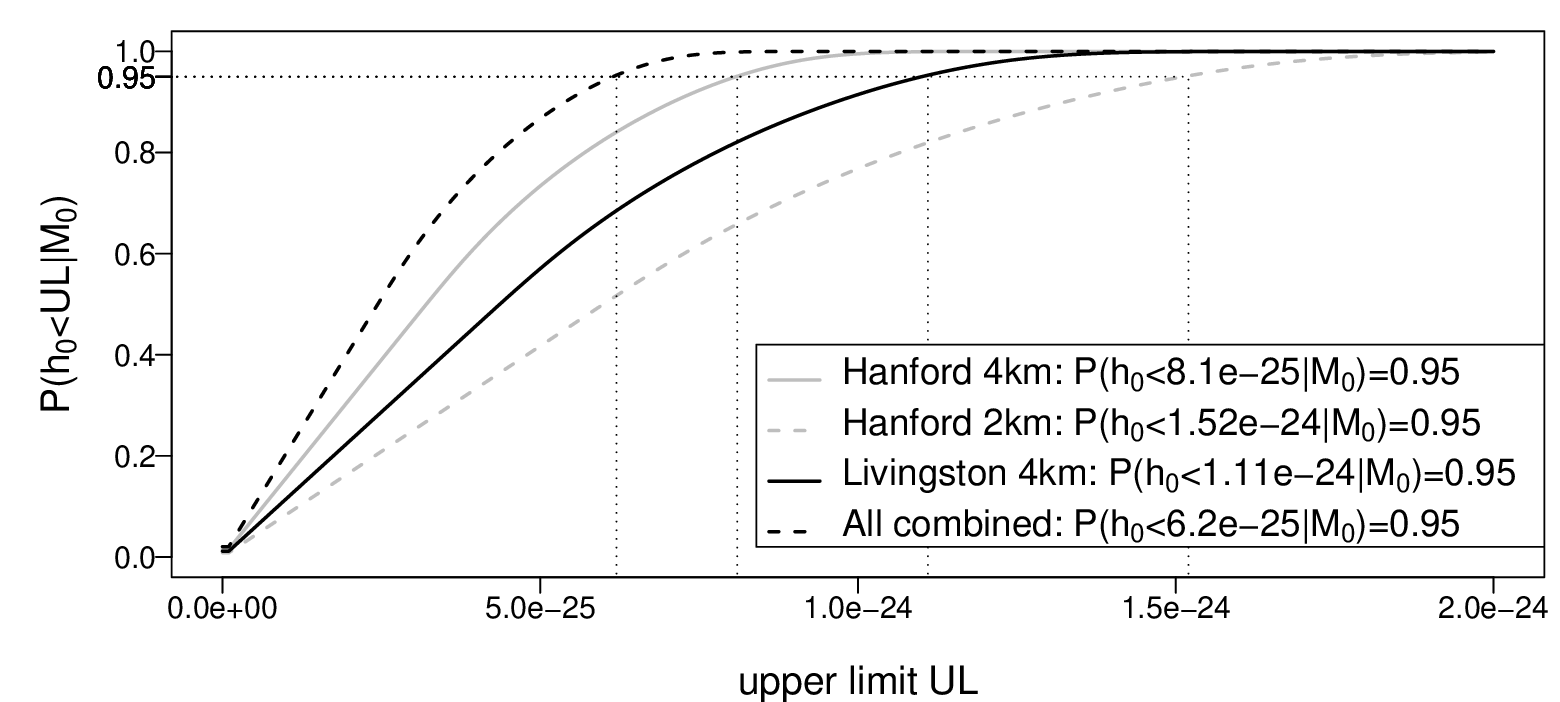}
  \end{center}
\caption{
Estimated sensitivity of the Bayesian method described in this paper, assuming 1 year of data with the typical noise level of the LIGO interferometers during their S5 run. The model prior probabilities are $\Prob(\mathcal{M}_1)=0.5$ (left) and  $\Prob(\mathcal{M}_1)=10^{-9}$ (right). The prior for $h_0$ is $h_0 \sim \textrm{Unif}(0,10^{-20})$. The assumed noise levels are $\sigma_{\textrm{\scriptsize H}_1}=8 \times 10^{-24}$, $\sigma_{\textrm{\scriptsize H}_2}=1.5 \times 10^{-23}$, and $\sigma_{\textrm{\scriptsize L}_1}=9 \times 10^{-24}$.}
\label{UL1}
\end{figure}

The upper limit estimate for a Bayesian MCMC search involves testing the hypothesis $h^{*}_0 < \UL$ vs. $h^{*}_0 \ge \UL$ under the assumption $\mathcal{M}_0$ that there is no signal in the data. The derivation of the probability $\Prob(h^{*}_0<\UL | \mathcal{M}_0)$ will shed light on this matter.
We condition on noise, observation period, and location and after integrating over the prior distributions of $\cos\iota^{*}$ and $\psi^{*}$, $\Prob(\mathcal{M}_0|h^{*}_0,\OP,\sigmavec,\alpha^{*},\delta^{*}) := \int{\int{\Prob(\mathcal{M}_0|\avecp{}) d\Prob_{\cos\iota}(\cos \iota^{*}) d\Prob_{\psi}(\psi^{*})}}$,  we obtain
\be
\Prob(h^{*}_0<\UL|\mathcal{M}_0,\OP,\sigmavec,\alpha^{*},\delta^{*})=\frac{\int_{0}^{\UL} \Prob(\mathcal{M}_0|h^{*}_0,\OP,\sigmavec,\alpha^{*},\delta^{*}) \Prob(h^{*}_0) d h^{*}_0 }
{\int_0^{\infty} \Prob(\mathcal{M}_0|h^{*}_0,\OP,\sigmavec,\alpha^{*},\delta^{*}) \Prob(h^{*}_0) d h^{*}_0}.
\label{upperlimit}
\ee
In order to derive Eq.~(\ref{upperlimit}) we need to find a suitable prior for $\Prob(h^{*}_0)$. One choice could be to put a uniform prior on $h_0$
with large boundary $[0,10^{-20}]$. The upper boundary of the prior range has negligible impact on the results of Eq.~(\ref{upperlimit}) as long as this boundary is significantly larger then the upper limit estimate. Fig.~\ref{UL1} displays Eq.~(\ref{upperlimit}) for two different prior probabilities on whether we expect a signal at SN1987a.

\begin{figure}[!ht]
  \begin{center}
    \includegraphics[width=0.48\textwidth]{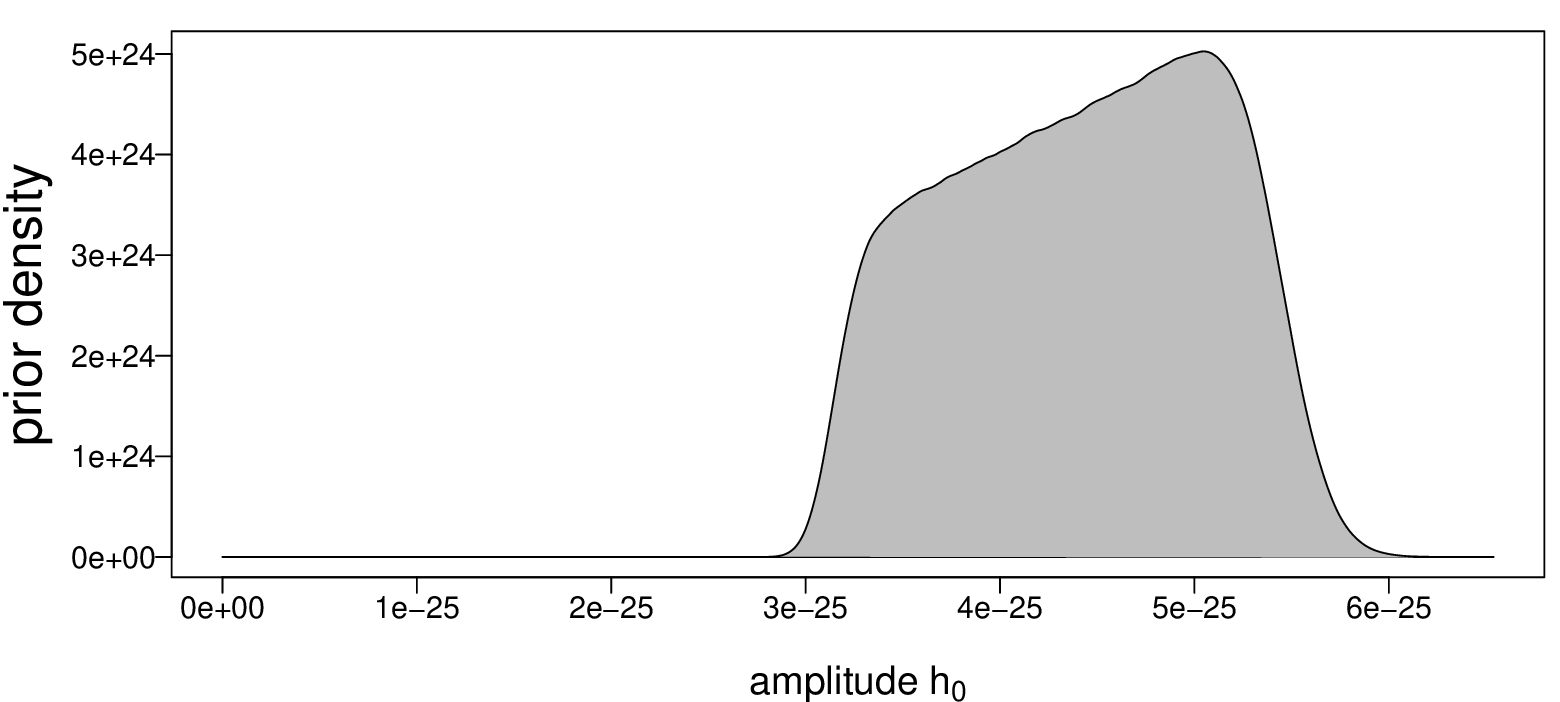}
    \includegraphics[width=0.48\textwidth]{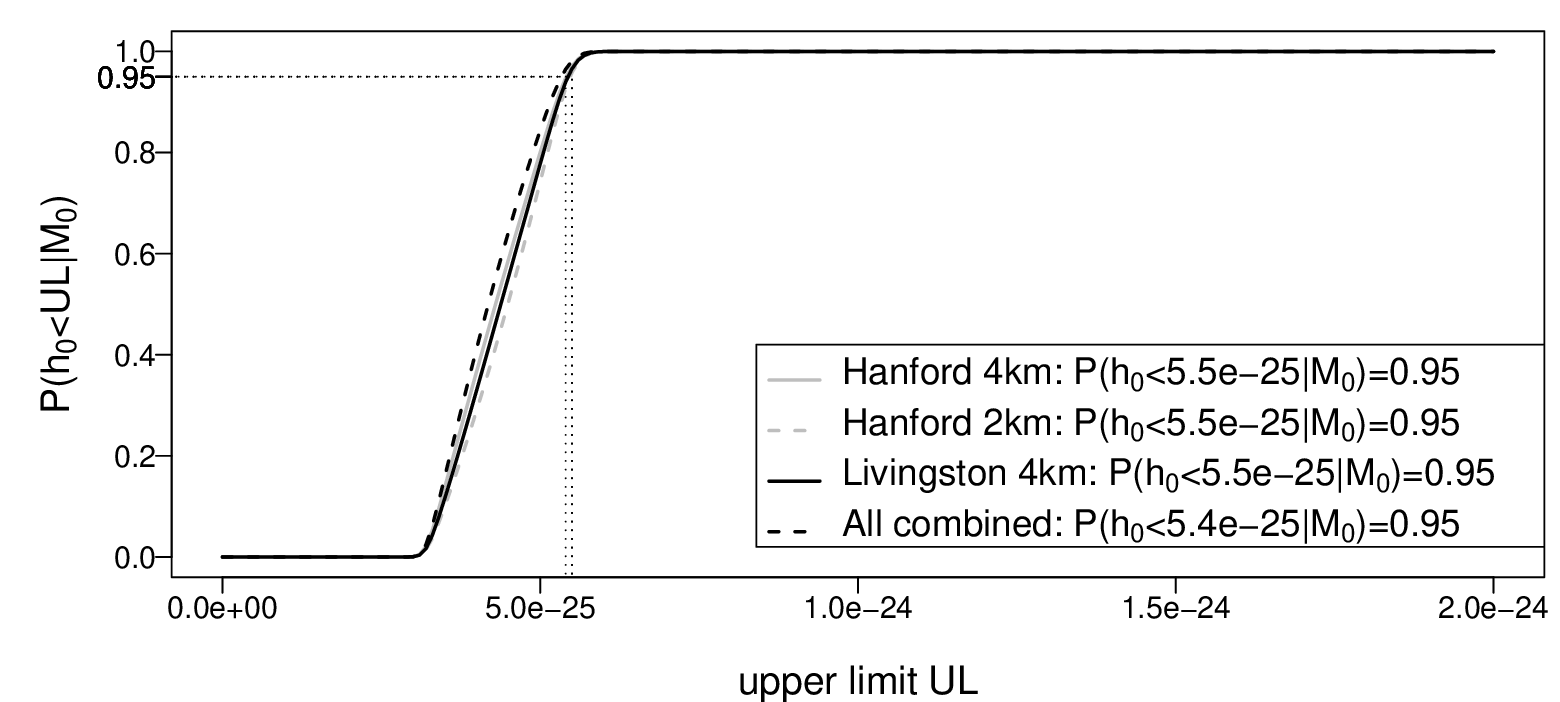}\\
    \includegraphics[width=0.48\textwidth]{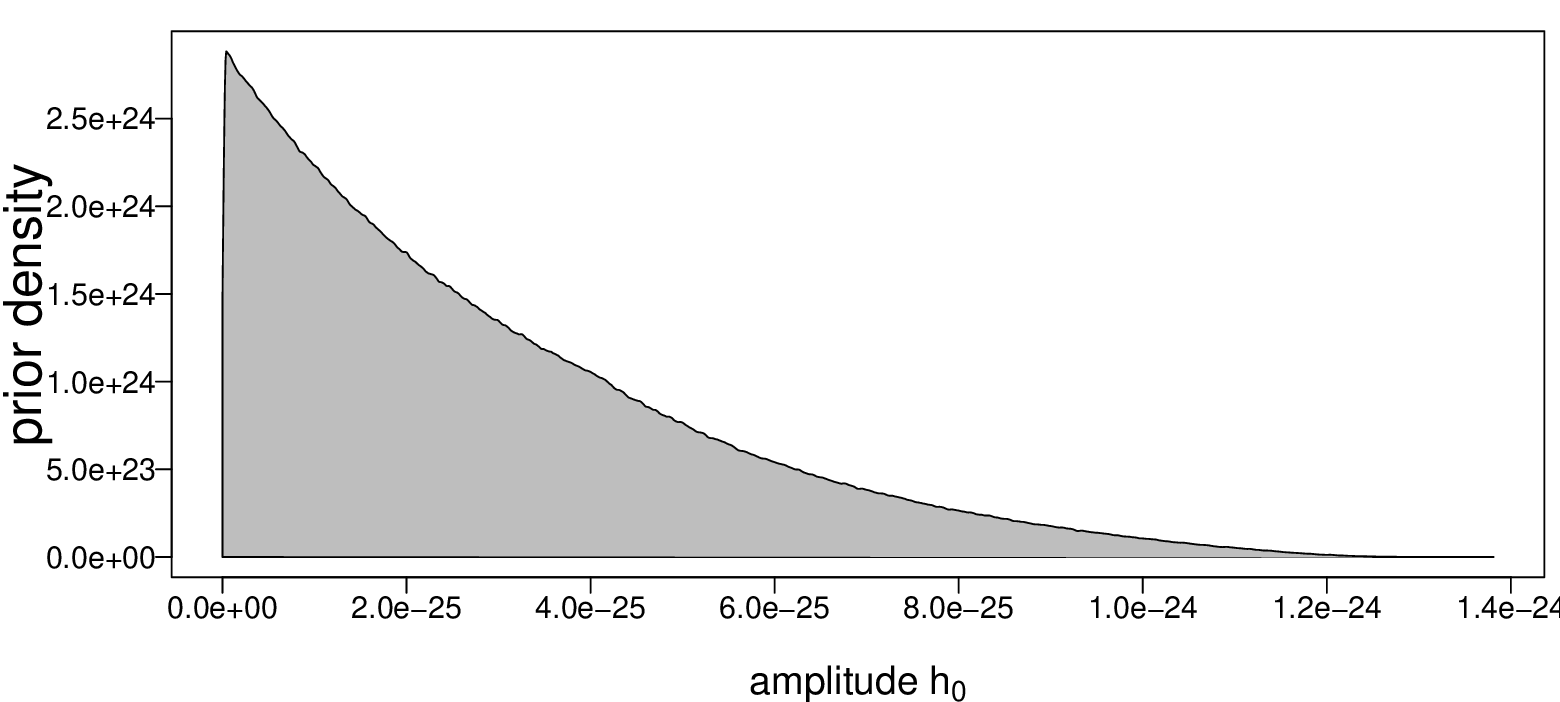}
    \includegraphics[width=0.48\textwidth]{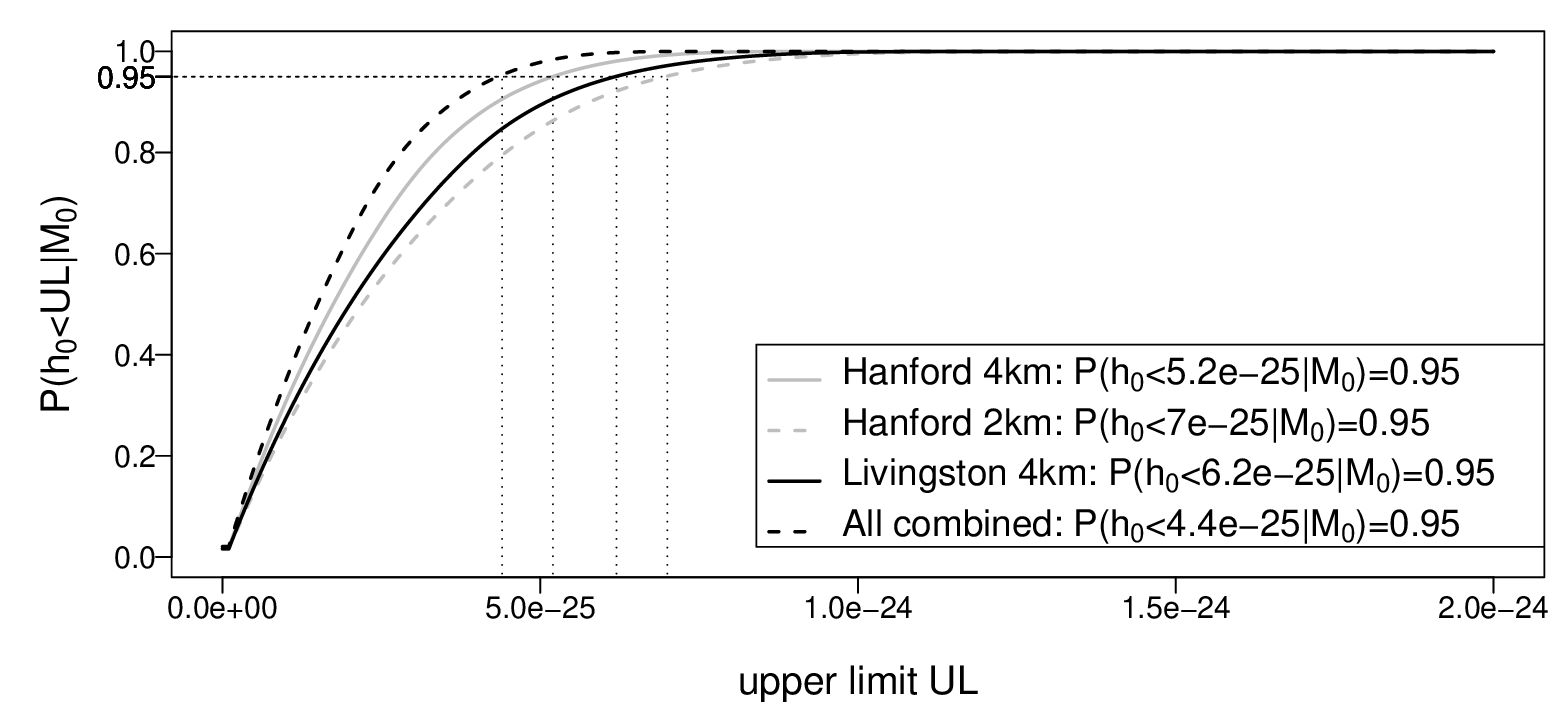}
  \end{center}
\caption{Two different prior distributions (left column) for $h_0$ for a possible neutron star in SN1987a and the corresponding curves for the Bayesian MCMC upper limit estimates (right column). 
The upper row corresponds to a prior subject to Eq.~(\ref{spindownmodel}) whereas the lower row is based on Eq.~(\ref{ellipticityfrequencymodel}). For the model selection, the prior probability is chosen to be $\Prob(\mathcal{M}_1)=10^{-9}$. }
\label{spindownANDclassic}
\end{figure}

Since we focus our search on a possible pulsar in SN1987a, we can tailor a prior distribution for $h_0$ as we know the age of SN1987a and its distance. In \cite{Abbott:2006} it is assumed that a newly formed neutron star spins at high rate and gravitational radiation slows it down. Two different prior scenarios are conceived here. According to \cite{Abbott:2006}, it is
\be
h_0(f)=r^{-1}\sqrt{(5GI_{zz})/(8c^3\tau_{\rm gw}(f)}),
\label{spindownmodel}
\ee
where $\tau_{\rm gw}(f)$ is the time for the gravitational wave frequency to drift down to frequency $f$ from its original spin rate. In case of SN1987a it is 20 years.
Here, $G$ is Newton's constant, $c$ the speed of light, $r$ the distance to the neutron star, and $I_{zz}$ the principal moment of inertia about the rotation axis.
In order to derive a prior distribution for $h_0$ we need to determine prior distributions for $r$ and $I_{zz}$.
We assume the distance estimated in \cite{1997AAS...191.1909P} with $50.9 \pm 1.8 $kpc  with $r \sim \rm{N}(50.9,1.8^2)$kpc for accounting the uncertainty in the distance. For the moment of inertia, we choose a uniform prior within the range $[10^{38},3\times 10^{38}]$ as applied in \cite{Abbott:2007}. These considerations yield a prior for $h_0$ as shown later in Fig.~\ref{spindownANDclassic}. 

A totally different approach for obtaining a prior distribution for $h_0$ is by \cite{LIGO-CW:2004}
\be
h_0=4 \pi^2 G I_{zz} f^2 \epsilon/(c^4 r)
\label{ellipticityfrequencymodel}
\ee
for a general pulsar expected at SN1987a. Here, $f$ is the pulsar's rotation frequency, and  $\epsilon$ its ellipticity.
We heterodyne to a frequency of $f_{\rm s}=2 f_{\rm r}=935$Hz, and assume the gravitational wave frequency to have this value within a $1/60$Hz frequency band for a particular search. An uncertainty beyond this needs to be accounted for in the prior $\Prob(\mathcal{M}_1)$ for the existence of the signal within the $1/60$Hz band around $f_{\rm s}$ because the signal is not seen outside that band after the heterodyning process.

We use the same uniform prior for $I_{zz}$ as above but we have to find a suitable prior for the ellipticity.
In \cite{Polomba:2005}, the ellipticity is assumed to have an exponential distribution (maximum entropy prior) with cut-off at a maximum ellipticity threshold.
Although in \cite{Polomba:2005} more pessimistic mean and maximum values are used, our choice is more optimistic in order to account for the fact that we know that a possible neutron star in SN1987a is very young.
We choose a cut-off according to \cite{Abbott:2006} at $\epsilon_{\max} \approx 9 \times 10 ^{-5}$ based on the idea of a hybrid neutron star with a mixed quark and baryon core and a normal neutron star in the outer part. For the mean of the exponential prior distribution we use an optimistic choice of $\epsilon_{\rm mean} = 5 \times 10^{-5}$.
Both prior distributions for $h_0$ as discussed above are displayed in Fig.~\ref{spindownANDclassic} along with their resulting upper limit estimates.

The use of such priors changes the results for upper limit estimates compared to those in Fig.~\ref{UL1} (which were based on a uniform prior).
In essence, a uniform prior on the amplitude recovers the detection ability of an interferometer. For example, in case of combined data sets, an upper limit estimate based on a uniform prior requires an amplitude of at least $6.2 \times 10^{-25}$.
The use of prior distributions based on Eq.~(\ref{spindownmodel}) and Eq.~(\ref{ellipticityfrequencymodel}), however, only have $0.001$\% and $11.4$\% probability mass above that limit, respectively. This inevitably yields values for the upper limits estimates dominated by the prior of $h_0$. This is obvious especially in case of the prior based on Eq.~(\ref{spindownmodel}) and can be seen in Fig.~\ref{spindownANDclassic}.

\section{\bf Conclusions}
The Bayesian MCMC methods work well when the SNR is sufficiently large but they struggle when the signal is too weak and the parameters that affect the phase evolution are not known. The fact that we integrate over very long observation
periods requires an almost exact match of the phase evolution and
almost all mass of the posterior distribution is highly concentrated around
one point in the parameter space when the SNR is large.
Finding this posterior peak with Bayesian MCMC methods is time consuming but once found, the sampling process is easy and efficient.
With decreasing SNR, however, the sampler is forced to also sample from other areas of the parameter space determined by the prior. This requires multiple retrievals of the narrow peak and it requires extremely long runs to gain insight into the actual shape of the posterior distribution.
The sampling speed depends on observation length and number of Markov chains involved when using parallel tempering. For one year of data, each single chain samples about $150\,000$ samples per week and chain on a 2.8 GHz machine. At low SNRs at least 10 chains are needed \cite{UmstaetterThesis:2006}. 
When no sensible inference can be drawn from an MCMC output if no frequency parameters can be retrieved. In those cases, the method derived here, based on model comparison, provides an excellent means for estimating an upper limit for the amplitude of a signal when using Bayesian MCMC methods, given the observation period and noise. In practice this method could be used to estimate the sensitivity of the Bayesian MCMC method on actual S5 data and in particular the noise. 
For long observation periods, the impact of prior information about the presence of a signal is rather small.
The influence of the amplitude's prior only becomes significant when the sensitivity, with respect to the obtained data, is too small for the expected amplitudes. 
Consequently, when we expect amplitudes below the detection limit then the upper limit estimate is determined mainly by the prior distribution of the amplitudes.
\ack
\small{This work was supported by the Marsden Fund Council from Government funding administered by the Royal Society of New Zealand (Grant UOA-204), and the National Science Foundation grant PHY-0553422.}

\section*{References}
  \bibliographystyle{unsrt}

\end{document}